# Influence of α-particles irradiation on the performance and defect levels structure of Al/SiO$_2$/p-type Si surface barrier detector

S.V. Bakhlanov[1], N. V. Bazlov[1,2], D.V. Danilov[2], A. V. Derbin[1], I. S. Drachnev[1], I. M. Kotina[1], O. I. Konkov[1,3], A.M. Kuzmichev[1], M.S. Mikulich[1], V. N. Muratova[1], M. V. Trushin[1], E. V. Unzhakov[1]

[1] NRC "Kurchatov Institute" - PNPI, Gatchina, Russia
[2] V.A. Fok Institute of Physics, Saint Petersburg State University, St. Petersburg, Russia
[3] Ioffe Physical-Technical Institute of the Russian Academy of Sciences, St. Petersburg, Russia

**Abstract.** Deterioration of the operation parameters of Al/SiO$_2$/p-type Si surface barrier detector upon irradiation with alpha-particles at room temperature was investigated. As a result of 40-days irradiation with a total fluence of $8\times10^9$ α-particles, an increase of α-peak FWHM from 70 keV to 100 keV was observed and explained by increase of the detector reverse current due to formation of a high concentration of near mid-gap defect levels. Performed CV measurements revealed the appearance of at least $6\times10^{12}$ cm$^{-3}$ radiation-induced acceptors at the depths where according to the TRIM simulations the highest concentration of vacancy-interstitial pairs was created by the incoming α-particles. The studies carried out by current-DLTS technique allowed to associate the observed increase of the acceptor concentration with the near mid-gap acceptor level at E$_V$+0.56 eV. This level can be apparently associated with V$_2$O defects recognized previously to be responsible for the space charge sign inversion in the irradiated n-type Si detectors.

## 1. Introduction

Unlike to the most of semiconductor electronic devices, the detrimental influence of nuclear radiation on the operational parameters of semiconductor detectors is inevitable during device application. Incoming radiation not only generates electron-hole pairs in the detector's sensitive area producing thus an information signal, but also creates radiation damage in the crystal lattice [1][2][3]. Accumulation of the radiation-induced defects reduces the nonequilibrium carrier lifetime and therefore degrades the main operational parameter of the detector – its energy resolution. Most effectively this process proceeds in case of irradiation by alpha particles, accelerated ions or fission fragments which are capable of transferring a significant fraction of their energy to the atoms of the detector lattice. As a result, already at a dose of $10^{10}$ cm$^{-2}$ a concentration of radiation defects reaches the value sufficient for a significant deterioration of the device performance.

Nowadays there is an increasing interest to the compact calibration neutron sources suitable for various research and industrial applications dealing with necessity to calibrate neutron measuring devices [4][5]. Possible solution could be a combination of $^{252}$Cf radionuclide with a semiconductor detector that detects fission fragments and thus provides a time reference of the neutron production. Silicon semiconductor detectors are capable of detecting fission fragments and α-particles due to their thin entrance window and high energy resolutions. Constraints on the usage of the semiconductor detectors are imposed by their finite radiation hardness.

This paper is devoted to the investigations of operational parameters degradation of surface barrier detector produced on p-type Si under irradiation with α-particles, which inevitably accompany spontaneous nuclear fission. The defects introduced during irradiation were studied with capacitance-voltage (CV) and deep level transient spectroscopy technique working in the current mode (IDLTS). The achievements in the understanding of the relation between the irradiation dose, degradation of the detector parameters and the created defect types and density would be essential for production of the effective radiation hard silicon detectors as well as for the optimization of possible calibration neutron sources.

## 2. Experimental Details

Detector for the experiments was fabricated from p-type boron-doped silicon wafer of (111) orientation. Diameter of the wafer was 10 mm, resistivity 2.5 kΩ×cm and a carrier lifetime 800 μs, respectively. After mechanical polishing the wafer was thoroughly cleaned and etched in $HNO_3$:HF solution. Afterwards, the wafer was oxidized in concentrated boiled $HNO_3$ solution in order to produced thin tunnel transparent oxide layer which serves as a passivation coating (so-called NAOS method [6]). Ohmic contact was made by evaporation of Pd layer on the whole rear side of the wafer, whereas the rectifying one – by evaporation of Al dot of 7 mm diameter in center of the wafer's front side.

Spectrometric channel for alpha-particles registration consists of charge-sensitive preamplifier connected with BUI-3K shaping amplifier with a shaping time of 2 μs and a 4000-channel 12-bit CAMAC analog-to-digital converter with 1.7 keV/ch resolution. Additionally, 200 Hz pulse signal was fed to the preamplifier to generate the "reference peak" enabling accurate separation of the intrinsic detector resolution from the noise contribution of the detector-amplifier system which are summarized in quadrature as

$$\sigma^2 = \sigma_{el}^2 + \sigma_{det}^2 \qquad (1)$$

where $\sigma_{det}$ – intrinsic dispersion of the detector (which accounts the contribution of the Fano factor, losses in the detector dead layer as well as the spread of α-particles energies from the radiation source) and $\sigma_{el}$ – contribution of electronics, which would define the full width at half-maximum (FWHM) of the reference peak. At a normal distribution, the energy resolution characterized by FWHM of a spectral line will be $2.355\sigma$ [7].

Detector was irradiated by a reference spectrometric ternary source containing $^{233}$U, $^{239}$Pu and $^{238}$Pu isotopes emitting α-particles at 4824 keV, 5156 keV and 5499 keV, respectively, with almost equal activities. The source represents a stainless steel substrate covered with a thin layer of the active material and the spread of α-particles energies from the source doesn't exceed 20 keV. For the measurements detector was mounted in a vacuum chamber operating at room temperature. Detector irradiation was performed through the collimator providing counting rate 2400 cps of α-particles. Reverse bias voltage applied to the detector was set to 10 V what was proved to be enough to register the α-particles (space charge region depth corresponding to 10 V was around 50 μm in our detector, while the penetration depth of α-particles in Si is about 30 μm) whereas the reverse current remained small.

For the traps characterization, SULA DLTS spectroscopy system working in a mode of current transients registration was used. Current DLTS (IDLTS) is well-suited for the measurements on high-resistive samples, where a capacitance DLTS encounters limitations as a high series resistance distorts the capacitive measurement. After routine assessment including current-voltage, capacitance-voltage (CV) and IDLTS characterization of the as-prepared detector, it was exposed to α–irradiation for 40 days up to the total fluence $\Phi$ of $8\times10^9$ α-particles, what corresponds to a dose of approximately $2\times10^{10}$ cm$^{-2}$. Acquisition of the detector reverse current was performed continuously during the whole irradiation period. Afterwards the irradiated detector was subjected to thorough spectroscopic and electrical characterization including IDLTS measurements.

## 3. Experimental Results

*3.1 Influence of α-irradiation on the detector performance*

The α-particle spectra measured by as-prepared detector and by the same detector after irradiation are shown in Fig. 1a, where along with three peaks due to α-particles of different energies the reference peak due to pulse signal are clearly visible. On the as-prepared detector the full width at the half maximum (FWHM) of 5499 keV peak was defined to be about 70 keV, whereas FWHM of the reference peak – 25 keV. After irradiation detector resolution has degraded – FWHM of 5499 keV peak has increased up to 100 keV and of the reference peak – up to 48 keV, respectively. However, the intrinsic dispersion of the detector $\sigma_{det}$ calculated by Eq. 1 remains quite similar before and after irradiation at about 27-28 keV. Thus it could be concluded, that the main contribution to the deterioration of the energy resolution of the irradiated detector is due to increase of $\sigma_{el}$ term in Eq. 1. The later includes the contributions of detector leakage current $\sigma_I$, detector capacitance $\sigma_C$ and feedback resistance $\sigma_R$ [7]

$$\sigma_{det}^2 = \sigma_I^2 + \sigma_C^2 + \sigma_R^2 \qquad (2)$$

Since the feedback resistance $R$ as well as the capacitance of the detector at the reverse bias of 10V remained unchanged during the irradiation (see Fig. 2a), the degradation of the detector performance is definitely caused by increase of the $\sigma_I$ term as the detector reverse current at applied reverse bias of 10V has increased from 0,2 µA up to 0,7 µA during the irradiation, see Fig. 1b. Measured current vs. fluence dependence can be satisfactorily approximated by the straight line with the slope of $\Delta I/\Delta\Phi \sim 6\times10^{-17}$ A/α (*sin*-like oscillation of the current visible in Fig. 1b are caused by the temperature instability during the irradiation process).

The dependence of $\sigma_I$ term in a circuit with a charge-sensitive preamplifier on the reverse current $I$ is determined by the charge fluctuations associated with the current and by the shaping time $\tau$ of the amplifier and could be expressed as

$$\sigma_I = k_I \sqrt{\tau I} \qquad (3)$$

The proportionality coefficient $k_I$ is defined by a specific shaping filter. In our case, the FWHM increase from 70 keV to 100 keV can be described by the increase of the reverse current from 0,2 µA up to 0,7 µA when the proportionality coefficient $k_I$ is equal to $3,3\times10^7$ keV/(A*sec)$^{1/2}$, as it is shown by the calculated dependence of peak FWHM on α-fluence in Fig. 1b. Obtained $k_I$ value is rather typical for the used preamplifier type [8]. According to Eq. 1-3 the linear increase of the reverse current should result in square root increase of the peaks FWHM, however for the considered range of fluences the FWHM dependence could be satisfactorily described by a simple linear dependence with the slope $\Delta\sigma/\Delta\Phi = 1.8\times10^{-9}$ keV/α, see Fig. 1b.

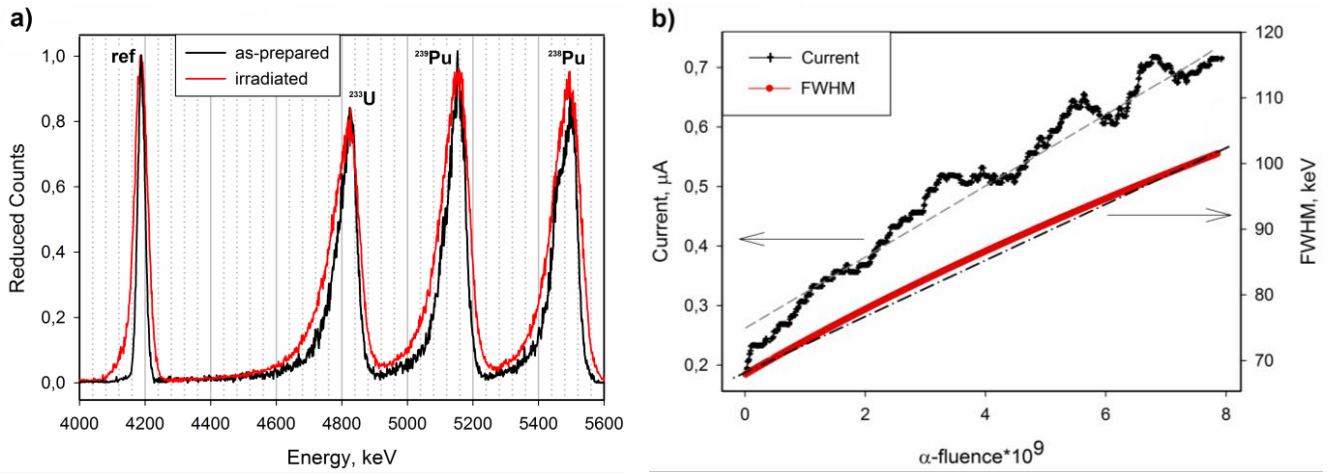

**Figure 1.** (a) The α-particle spectrum in the range of 4600–5600 keV and the 'reference' peak at 4200 keV measured by as-prepared and by irradiated detector; (b) increase of the detector reverse current at applied bias 10V and the calculated growth of the α-peak FWHM with the fluence. Thin dashed line shows the best linear fit to the current dependence, whereas dash-dotted one – to that of FWHM.

*3.2 CV-profiling*

On the as-prepared detector DLTS measurements didn't reveal any deep levels in the active region of the detector where the absorption of α-particles takes place, whereas CV profiling revealed flat distribution of the acceptor concentration at the level of $4.5\times10^{12}$ cm$^{-3}$ what corresponds to the wafer resistivity. After irradiation, a local increase of the detector capacitance in the reverse bias range of 1-8 V was detected by the CV measurements, see Fig. 2a (note, that the capacitance value at the reverse bias of 10V which was kept during the detector irradiation remained unchanged). And at the corresponding profile of the ionised acceptor distribution shown in Fig. 2b a broad peak reaching the value of $1.1\times10^{13}$ cm$^{-3}$ appeared which was certainly absent in the as-prepared detector.

In order to define possible correlation between the peak on acceptor profile with the distribution of primary radiation defects (vacancies and interstitials) generated during α-irradiation, the numerical simulation of the vacancies distribution in silicon matrix as a result of irradiation by α-particles with three energies (4824 keV, 5155 keV and 5499 keV) was performed using TRIM program [9]. Calculated vacancy profiles are also shown in Fig. 2b together with the ionized acceptor distributions in the as-prepared and the irradiated detector. As one can see in Fig. 2b, peak on acceptor profile appears approximately at the same depth where the concentrations of vacancies generated by α-particles reach their maxima. Note that TRIM doesn't account for any possible vacancy interaction in the solid (including vacancy-interstitial recombination), so the direct comparison of vacancies and radiation-induced acceptor concentrations in Fig. 2b has no sense.

Generally, the abrupt step in the ionized charge distribution occurring at the distances less than Debye length $L_D$ (1.5-2 μm for the doping of our sample) could not be faithfully reproduced by CV profiling. The result of the profile calculation from CV curve in case of such abrupt charge variation would be a profile "smeared" by approximately $2L_D$ (i.e. by 3-4 μm in our case) on both sides of a real charge step [10]. This explains the minor deviation of the calculated acceptor profile from the profiles of vacancies distribution in Fig. 2b, namely that the acceptor concentration starts to growth from the depth of 35 μm where according to TRIM simulation vacancy concentration should be zero and that the acceptor profile maximum is slightly shifted towards the shallower depths relative to the integrated vacancy profile maximum. Thus it could be concluded that the local increase of the acceptor concentration in the irradiated detector is caused by the creation of the negatively-charged radiation-induced defects as a result of α-particles irradiation. Considering the difference between the ionized acceptor profiles in the as-prepared and irradiated detector, the density of such radiation-induced acceptor-like defects could be estimated to be around $N_{RA}=6\times10^{12}$ cm$^{-3}$.

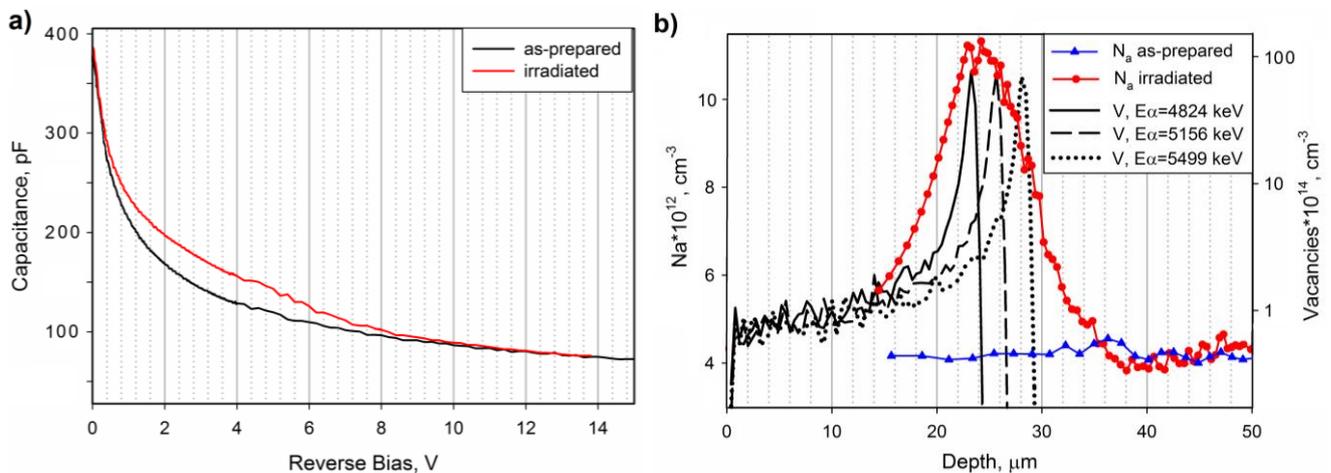

**Figure 2.** (a) CV curves measured on as-prepared and irradiated detector at room temperature; (b) Corresponding acceptor distribution profiles in the as-prepared and irradiated detector together with the vacancies distributions calculated by TRIM program for α-particles of three different energies.

*3.3 IDLTS results*

To define the parameters of defect levels introduced by α-irradiation into the Si band gap, current DLTS (IDLTS) measurements were performed. Concentrations of the detected traps were recalculated by using the expression describing the magnitude of IDLTS signal

$$N_t = \frac{2S(T)t}{Akq(w_r - w_p)}, \qquad (4)$$

where $S(T)$ is the temperature dependent IDLTS signal (in ampere), $t$ is the rate window period, $k$ is the proportionality coefficient which depends on exact correlator system (for double box-car correlator $k=0.363$) and $w_r$ and $w_p$ are the space charge region (SCR) depths corresponding to the applied reverse bias $U_r$ and the pulse voltage $U_p$ used for the measurements [10]. For IDLTS peaks appearing at different temperatures the respective SCR depths were recalculated from capacitance values measured at corresponding temperatures.

Three IDLTS spectra were measured to study the defect distribution at various depths below the surface. The first spectrum was recorded with the reverse bias voltage $U_r$=5V and the pulse voltage $U_p$=1V, what corresponds to space charge region depth variation from $w_r$=30 μm to $w_p$=18 μm, respectively. Thus, the probing depth for this spectrum just corresponds to the depth where the acceptor peak appeared (see Fig. 2b). Two peaks appearing nearly 205K and 280K of considerably different magnitudes could be clearly resolved in the spectrum, see Fig. 3a. Concentrations of corresponding traps as calculated by Eq. 4 were found to be around $6 \times 10^{11}$ cm$^{-3}$ and $3 \times 10^{12}$ cm$^{-3}$. The activation enthalpies $E_a$ and capture cross sections defined from the Arrhenius plots (Fig. 3b) are 0.56 eV and $6 \times 10^{-14}$ cm$^2$ for deeper traps and 0.38 eV and $5 \times 10^{-14}$ cm$^2$ for shallower ones, respectively.

At another spectrum recorded with $U_r$=10V and $U_p$=5V, i.e. when SCR depth varied from $w_r$=50 μm to $w_p$=30 μm – what is definitely deeper than the depth of the acceptor peak and the depths of the vacancies distribution maxima in Fig. 2b – only the peak due to deeper traps at around 280K remains visible with its magnitude decreasing drastically by nearly one order of magnitude as compared to the previous spectrum.

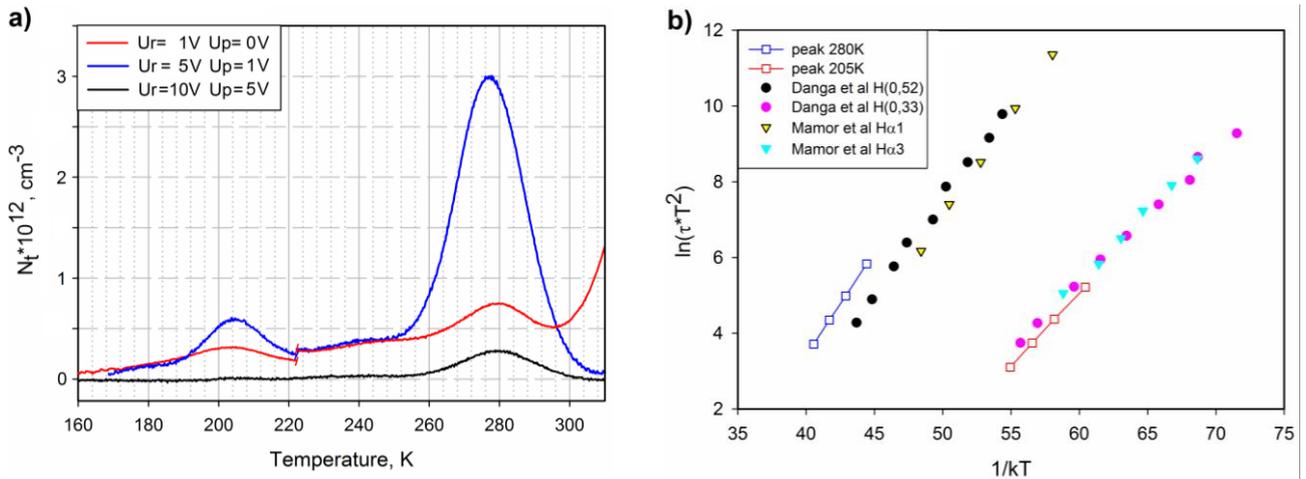

**Figure 3.** (a) IDLTS spectra recorded with different voltage parameters (see comments in the text), filling pulse duration and rate window period were 1 ms. The part of the spectra below 220K were recalculated to represent trap concentration $N_t$ using capacitance values measured at 205K, whereas above 220K – measured at room temperature, respectively. (b) Comparison of the Arrhenius plots for the traps revealed in the irradiated detector (empty squares) with those ones reported previously in [11] (circles) and [12] (triangles) for the deep traps in p-type Si samples irradiated by α-particles.

The third spectrum recorded with $U_r=1$V and $U_p=0$V ($w_r=18$ μm and $w_p=10$ μm), i.e. when the probing depth was closer to the surface than the acceptor peak in Fig. 2b, has revealed both 210K and 280K peaks with their magnitudes being considerably lower than in the first spectrum recorded with $U_r =5$V and $U_p=1$V. Additional signal visible above 300K is coming due to surface states at Si/SiO$_2$ interface and was presented in the spectrum before irradiation. Lower traps concentrations revealed by this spectrum agree with the lower concentration of radiation induced vacancies expected at such depths as it follows from the TRIM simulations (Fig. 2b).

## 4. Discussion

As a result of the prolonged room temperature irradiation of p-type Si surface barrier detector with α-particles up to a total fluence of $8 \times 10^9$, a degradation of the detector energy resolution from 70 keV to 100 keV of α-peak FWHM was observed. Increase of α-peaks FWHM was related with irradiation-induced increase of the reverse current which via $\sigma_I$ dispersion term (Eq. 3) influences the detector resolution. The increase of the reverse current with α-fluence was found to proceed nearly linearly with the slope of $\Delta I/\Delta\Phi \sim 6\times10^{-17}$ A/α, what in turn results in a square root dependence of α–peak FWHM on the fluence, which however could be satisfactorily described by the linear dependence with the slope of $\Delta\sigma/\Delta\Phi = 1.8\times10^{-9}$ keV/α for the used detector and pre-amplifier. Thus we may predict the α-peak FWHM at a fluence of $\Phi = 10^{10}$ to be around 112 keV. Such peak broadening will not prevent the reliable distinction between the α-peaks of different energies in Fig. 1a as well as between the peaks due to α-particles and fission fragments when using such detector as a part of calibration neutron source. The increase of the detector reverse current and hence the increase of α-peak FWHM could be diminished by cooling the detector. Recently the slope values of $\Delta I/\Delta\Phi \sim 1.4\times10^{-17}$ A/α and $\Delta\sigma/\Delta\Phi \sim 8.4\times10^{-10}$ keV/α, respectively, were obtained during prolonged irradiation of Si(Li) p-i-n detector by the same α-source at liquid nitrogen temperature [13]. Comparing with the slope coefficient obtained in this work it follows that fourfold decrease of the current growth coefficient $\Delta I/\Delta\Phi$ results in twofold decrease of the peak FWHM broadening coefficient $\Delta\sigma/\Delta\Phi$.

By CV measurements a local increase of the acceptor concentration up to a value of $1.1\times10^{-13}$ cm$^{-3}$ in the region of highest vacancy-interstitial pairs generation by incident α-particles relative to the doping level of $4.5\times10^{12}$ cm$^{-3}$ was revealed after irradiation. And by performing current-DLTS measurements two deep levels reaching the highest concentration just at the same depth as the acceptor peak on CV profile of irradiated detector (Fig. 2b) were detected. Arrhenius plots for these two peaks were compared in Fig. 3b with those ones reported previously for the deep traps observed in α-irradiated p-type Si samples. Close coincidence was established between the Arrhenius graph for 205K peak with those for H(0.33) [11], Hα3 [12] and H(0.35) [14] (not shown in Fig. 3b) peaks, respectively, which were attributed to the C$_i$-related defects. The charge state of these defects was found to be 0/+, i.e. zero when traps are empty and positively charged when filled with holes. Thus, this defect level couldn't be responsible for the increase of the acceptor concentration in the irradiated detector. Moreover, the concentration of such traps as defined by IDLTS is an order of magnitude lower than $N_{RA}$ concentration of radiation-induced acceptors.

On the other side, the concentration of the deeper trap of about $3\times10^{12}$ cm$^{-3}$ is close to $N_{RA}$ value of $6\times10^{12}$ cm$^{-3}$ defined from CV measurements, what allows us to suppose that just this trap is responsible for the additional acceptor charge in the irradiated detector – especially by taking into account that IDLTS method underestimates the concentration of near mid-gap levels when the emission rates towards the valence and conduction band are comparable [10]. Previously, a deep level lying at 0,52 eV above the valence band edge $E_V$ was observed in α-irradiated p-type Si [11][12]. However, the Arrhenius plots for our 280K peak and for the corresponding H(0.52) [11] or Hα1 [12] peaks seem not to be identical, see Fig. 3b, and the defect level in our sample lies somewhat deeper in the band gap at $E_V+0,56$ eV. The exact origin of H(0.52) or Hα1 defects as well as their charge state were not established, whereas the doping level of irradiated p-Si samples in

[11][12] of a few $10^{15}$ cm$^{-3}$ made hardly detectable any possible changes in the acceptor concentration at a level of a few $10^{13}$ cm$^{-3}$ on CV profiles.

For the $E_V$+0.56 eV traps revealed in our detector the following possible origin could be supposed. There is a well-known phenomenon of so-called "space charge sign inversion" observed for Si detectors produced on n-type wafers – what means gradually inversion of initial n-type conductivity into p-type one with the increase of the irradiation dose of gammas, protons or pions (which similar to alphas introduce mainly point-like defects) [1][2][3]. This phenomenon was explained by the creation of a large number of deep radiation-induced acceptor defects whose negative charge first compensates the positive charge of the donor doping impurity in n-Si, whereas further irradiation accompanying with subsequent growth of the radiation-induced acceptor concentration leads to a prevalence of the negative charge in the space charge region and to conversion of the material into p-type. As the acceptor defects responsible for charge inversion the so-called *I* defects (which were tentatively ascribed to the divacancy-oxygen complex $V_2O$) were suggested [15] with their level lying almost at the middle of Si band gap at 0.55 eV below the conduction band edge. These defects were shown to be predominantly generated in FZ silicon, while their production is largely suppressed in oxygen-reach CZ silicon due to competitive defect reaction leading to the creation of vacancy-oxygen VO complexes [15].

Moreover, in the work of I. Pentilie et al [15] was shown that the level of *I* defect may exchange by charge carriers with both conductivity and valence band, what in DLTS experiments results in the appearance of two peaks with different activation energies of 0,55 eV for the case of electron emission towards the conductivity band and 0,58 eV for holes emission towards the valence band. The later value is pretty close to the activation energy of 0.56 eV derived for our DLTS peak appearing near 280K. Thus we may assume that in our detector (also produced on FZ silicon) the radiation-induced increase of the acceptor concentration is associated with the deep defects giving rise to DLTS peak at 280K, which are of the same nature as those ones responsible for conductivity type inversion in n-type silicon detectors under gamma (proton, pion) irradiation. It could be also suppose that in high-resistivity p-type Si as in our case, when the Fermi level lies near the middle of the band gap, this defect level would be empty of holes, i.e. it would poses a negative charge. Furthermore, the near mid-gap level of *I* (or $V_2O$) defect was recognized previously as giving the main contribution to increase of the detector reverse current under irradiation [1]. The same would be certainly true for our detector as well, where exactly the near mid-gap level at $E_V$+0.56 eV contributes to the leakage current growth.

**5. Conclusions**

The Al/SiO$_2$/p-type Si surface barrier detector was subjected to prolonged room temperature irradiation by α-particles produced by decay of $^{233}$U, $^{238}$Pu, and $^{239}$Pu nuclei with a total fluence of $8\times10^9$ α-particles. After such treatment, the degradation of the detector energy resolution from 70 keV to 100 keV of α-peak FWHM was revealed and related with increase of the detector reverse current which proceeds linearly with α-fluence with the slope of $\Delta I/\Delta\Phi \sim 6\times10^{-17}$ A/α. Accounting for such current growth the changes in the energy resolution could be satisfactorily described by linear dependence of the dispersion σ on α-fluence with the slope $\Delta\sigma/\Delta\Phi = 1.8\times10^{-9}$ keV/α for the particular spectrometric setup. Nevertheless, such degradation of the detector energy resolution would not prevent the reliable separation of the signals from α-particles emitting by such ternary α-source and in case of similar detector use for fission fragments detection – the separation of the α-particles signal from the signal due to the fission fragments up to α-fluence of $10^{10}$ at least.

Leakage current increase in the irradiated detector was related with the appearance of the radiation-induced defect level at $E_V$+0,56 eV whose depth distribution correlates with that for vacancy-interstitial pairs produced by the incoming α-particles. This level was found to introduce an additional acceptor-like charge in the detector active region up to a level of $N_{RA} \sim 6\times10^{12}$ cm$^{-3}$. We suppose that this level may be identified with $V_2O$ defects detected previously in irradiated FZ n-type silicon where they were considered to be responsible for space charge sign inversion from initial n-type into p-type upon irradiation by introducing an acceptor level in the middle of Si bandgap.

**Acknowledgments**
The reported study was funded by RFBR, project number 20-02-00571